\documentstyle[twoside,fleqn,espcrc2,epsfig]{article}
\newcommand{\be}{\begin{equation}}
\newcommand{\ee}{\end{equation}}
\newcommand{\bea}{\begin{eqnarray}}
\newcommand{\eea}{\end{eqnarray}}
\newcommand{\f}{\frac}
\newcommand{\Di}{{\cal D}}
\newcommand{\condensate}{\langle\bar{\psi}\psi\rangle}

\title{Spectrum of the fixed point Dirac operator in 
the Schwinger model\thanks{Presented by F. Farchioni; supported by Fonds 
zur F\"orderung der Wissen\-schaft\-lichen Forschung 
in \"Osterreich, Project P11502-PHY.}}

\author{F. Farchioni, I. Hip, C.~B. Lang 
and M. Wohlgenannt\\
\vspace{3mm}
Institut f\"ur theoretische Physik, 
               Universit\"at Graz,\\
               Universit\"atsplatz 5, A-8010 Graz, Austria}
       
\begin{document}
\pagestyle{empty}

\begin{abstract}
Recently, properties of the fixed point action for fermion
theories have been pointed out indicating realization 
of chiral symmetry on the lattice. 
We check these properties by numerical
analysis of the spectrum of a parametrized fixed point Dirac
operator investigating also microscopic fluctuations
and fermion condensation.
\end{abstract}

\maketitle

\section{INTRODUCTION}

Ginsparg and Wilson \cite{GiWi82} provided
an explicit condition (GWC) for the realization of the chiral symmetry 
on the lattice,
\be
\frac{1}{2}\left\{\, {\cal D},\gamma^5\,\right\}=
{\cal D}\,\gamma^5\,R\,{\cal D}\;\;,
\label{eq:gwcg}
\ee
where  ${\cal D}$ is the lattice Dirac operator and $R$ is an operator acting 
on space-time and color indices; the ${\cal O}(a)$
violation of the symmetry is local since $R$ is.
The GWC ensures the restoration of the main properties of the continuum theory 
related to chirality \cite{GiWi82,Ha98c,HaLaNi98,Ha98a,Lu98}. 

Recently it has been realized \cite{Ha98c} that the fixed point (FP) 
action of a block-spin transformation (BST) is a solution of the GWC.
An independent solution comes from the overlap formulation of chiral fermions, 
Neuberger's lattice Dirac operator \cite{Ne98}, 
obtained by a projection of the original Wilson operator 
with negative quark mass. 

Here we check the expected classical and quantum chiral features of 
the FP Dirac operator of the Schwinger model 
in a parametrized form \cite{LaPa98b}.
The cut-off of the less localized couplings in the parametrization is expected
to introduce deviations from the ideal behavior.
Some of the results presented  here appeared already 
(with lower statistics) in \cite{FaLaWo98}.

\section{GINSPARG-WILSON FERMIONS}

For a non-overlapping BST, the operator $R$ in 
(\ref{eq:gwcg}) is diagonal: $R=\f{1}{2}$, and (\ref{eq:gwcg}) 
assumes the form (shared also by Neuberger's operator)
\be
{\cal D}\,+\,{\cal D}^{\dagger}\:=\:{\cal D}^{\dagger}\,{\cal D}\:=\:{\cal D}\,{\cal D}^{\dagger}\;\;.
\label{eq:gwc}
\ee

As a consequence of (\ref{eq:gwc}), ${\cal D}$ is a normal operator,
its spectrum lies on the circle $|\lambda-1|=1$ in the complex plane
and all real modes have definite chirality. Moreover a `lattice' Index
Theorem (IT) holds \cite{HaLaNi98}:
\be
Q_{FP}\:=\:\sum_{\{v_0\}}\:(v_0,\gamma^5 v_0)\;\;,
\label{eq:AST}
\ee
where $Q_{FP}$ is the FP  topological charge of the background
gauge configuration and ${\{v_0\}}$ the set of zero modes of $\Di$. 

The GWC allows the definition of a subtracted fermion condensate
\cite{Ne98a,Ha98a} representing (for the number of fermion species
$N_f>1$) an order parameter for the spontaneous breaking of the
symmetry,
\be
\condensate_{\rm sub}\:=\:-\f{1}{V}
\left\langle\,{\rm tr}({\tilde\Di}^{-1})\right\rangle_{\rm gauge}\;.
\label{eq:subfc}
\ee
Because of (\ref{eq:gwc}), the redefined fermion matrix
$\tilde{\Di}=D\,(1-D/2)^{-1}$ is anti-hermitian and so its spectrum
$\{\tilde{\lambda}\}$ is purely imaginary. The corresponding spectral
density $\tilde{\rho}(\tilde{\lambda})=1/V(dN/d\tilde{\lambda})$
complies with the Banks-Casher formula:
\be\label{BanksCasher}
\condensate_{\rm sub}\:=\:-\pi\,\tilde{\rho}(0)\;\;.
\ee

$\tilde{\Di}$ anti-commutes with $\gamma_5$ and so the microscopic
fluctuations of its spectrum are expected \cite{Ve94} to be described
by the chiral Random Matrix Theory (chRMT) which in the case of
irreducible matrices predicts just three classes of universality
corresponding to orthogonal, unitary or symplectic ensembles
distributed according to the Gaussian measure (chGOE, chGUE chGSE
respectively).  The universal behavior can be probed \cite{aavv98}
through the microscopic spectral density $\rho_m(z)$
\be
\rho_m(z)\:=\:\lim_{V\rightarrow\infty}\f{1}{\Sigma}\,
\tilde{\rho}\left (\f{z}{V\Sigma}\right)\;,
\ee
(with $\Sigma=-\condensate$) and the distribution of the 
smallest eigenvalue $P(z_{\rm min})$.

\section{RESULTS}

The parametrized (fermion) FP action considered  here \cite{LaPa98b}
has the usual form in terms of paths (and loops) of link variables; in
our case these are enclosed in a $7\times7$ lattice, with altogether
429 terms per site. It was determined for non-compact gauge fields. For
the present context we studied the associated Dirac operator $\Di_p$
for samples of $5000$ configurations on a $16^2$ lattice, generated
according to both compact and non-compact standard Wilson gauge action
(quenched generation) for $\beta=2$, 4, 6 (for reason of space we
report the results mostly for the compact case).  The unquenching has
been realized by properly including the fermion determinant in the
observables.

\begin{figure}[t]
\begin{center}
\vspace*{-1mm}
\epsfig{file=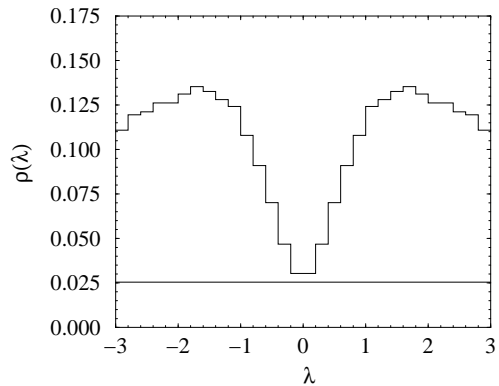,angle=-90,width=6cm}
\vspace*{-3mm}
\end{center}
\caption{\label{fig:dist}
Spectral density; $\beta=4$, lattice size $16^2$. The horizontal line
gives the theoretical expectation from the Banks-Casher formula.}
\end{figure}

\subsection{Spectrum}

The spectrum of $\Di_p$ is close to circular, somewhat fuzzy at small
values of $\beta\leq 2$ but excellently living up to the theoretical
expectations at large gauge couplings $\beta\geq 4$. The dispersion
$\sigma$  of the eigenvalues around the circle follows the law $\sigma$
$\propto1/\beta^{2.41}\simeq a^5$

Already at low values of $\beta$ there is a clear distinction between
real modes accumulating around 0 and around 2 (ideally 0 and 2 are the only possible real values).  We checked
their chirality finding a distribution sharply
peaked around $-1$ and 1 for
$\beta\geq 4$. Each (quasi) zero mode has either three ($20 \%$ of
cases at $\beta=6$) or one companion, the overall chirality of real
modes being zero.

We checked the IT adopting the geometric definition for the topological
charge of the background gauge configuration (in principle one should
use the FP topological charge: the two definitions agree for smooth
configurations).  We obtain $97 \%$ of successes already at $\beta=2$
and $100 \%$ for $\beta\geq 4$.

\subsection{Fermion condensate}

The infinite volume fermion condensate $\condensate$ can be obtained
from the spectral density $\tilde{\rho}(\tilde{\lambda})$ through the
Banks-Casher formula (\ref{BanksCasher}).  In Fig. \ref{fig:dist} we
report our results for the spectral density
$\tilde{\rho}(\tilde{\lambda})$ comparing to the value of
$\tilde{\rho}(0)$ expected according to $\condensate$ of the
continuum.

We also determined the finite volume condensate $\condensate_V$ by use
of the direct definition (\ref{eq:subfc}), obtaining (lattice units)
0.072(7) for $\beta=4$ and 0.063(3) for $\beta=6$ (continuum result for
the corresponding physical volume \cite{SaWi92}:  0.080 and 0.065).

\subsection{Microscopic fluctuations}

\begin{figure}[t]
\begin{center}
\epsfig{file=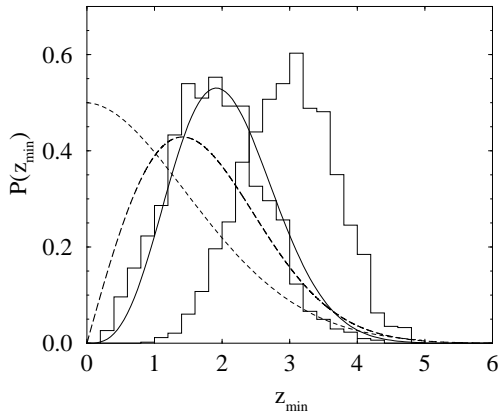,angle=-90,width=6.5cm}
\vspace*{-3mm}
\end{center}
\caption{\label{fig:minev}
Distribution of the smallest eigenvalue $P(z_{\rm min})$: $\beta=4$, 
lattice size $16^2$, $\nu=0$ (thick full lines) and $|\nu|=1$.
We compare with the predictions of the chRMT:
chGOE (dotted line), chGUE (dashed line), chGSE (full line).}
\end{figure}

We report here results just for the quenched ($N_f=0$) situation.  In
Fig. \ref{fig:minev} we present our outcomes for $P(z_{\rm min})$ for
the $\nu=0$ and $|\nu|=1$  topological sectors (the topological charge
has been counted according to the number of zero modes of $\Di_p$).
For the trivial sector we can compare with the three variants
\cite{Fo93} predicted by the chRMT: the chGSE seems to give the best
agreement. We also checked different lattices and values of $\beta$
finding consistency with the expected universality.  The results for
the non-compact gauge ensemble, $\nu=0$, agree within the statistical
errors.  The topological excitations seem to produce just a shift of the
distribution by $|\nu|$.

\begin{figure}[t]
\begin{center}
\epsfig{file=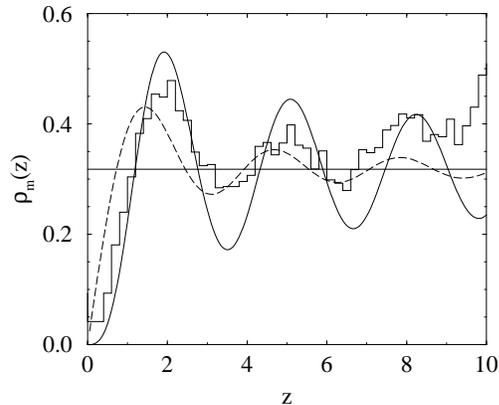,angle=-90,width=6.5cm}
\vspace*{-3mm}
\end{center}
\caption{\label{fig:micro}
Microscopic distribution $\rho_m(z)$: $\beta=4$, 
lattice size $16^2$ and $\nu=0$, and predictions of the chRMT,
chGUE (dashed line) and  chGSE (full line). The straight line
represents the asymptote for $\rho_m(z)$ ($1/\pi$).}
\end{figure}

Fig. \ref{fig:micro} reports the microscopic spectral density
$\rho_m(z)$ for the non-compact gauge field ensemble 
(in the $\nu=0$ sector). Here we find
disagreement with all three predictions of chRMT
\cite{Ve94bVeZa93NaFo95} indicating possible problems with some of the
assumptions (as maybe irreducibility of the matrices).

We thank P. Hasenfratz, A. Jackson and K. Splittorff for discussions.

\end{document}